\documentclass[twocolumn, dvipdfmx]{aastex631}
\usepackage{booktabs,}
\usepackage{multirow}
\usepackage{ulem}

\shorttitle{Soft X-ray Energy Spectra in the Galactic Disk with HaloSat}
\shortauthors{Ampuku et al.}
\graphicspath{{./}{figures/}}

\begin{document}

\title{Soft X-ray Energy Spectra in the Wide-Field Galactic Disk Area Revealed with HaloSat}

\correspondingauthor{Ikuyuki Mitsuishi}
\email{mitsuisi@u.phys.nagoya-u.ac.jp}

\author{Kazuki Ampuku}
\affiliation{Graduate School of Science, Division of Particle and Astrophysical Science, Nagoya University, Furo-cho, Chikusa-ku, Nagoya, Aichi, 464-8602, Japan}

\author{Ikuyuki Mitsuishi}
\affiliation{Graduate School of Science, Division of Particle and Astrophysical Science, Nagoya University, Furo-cho, Chikusa-ku, Nagoya, Aichi, 464-8602, Japan}

\author{Koki Sakuta}
\affiliation{Graduate School of Science, Division of Particle and Astrophysical Science, Nagoya University, Furo-cho, Chikusa-ku, Nagoya, Aichi, 464-8602, Japan}

\author{Philip Kaaret}
\affiliation{NASA/Marshall Space Flight Center, Huntsville, AL 35812}
\affiliation{University of Iowa Department of Physics and Astronomy, Van Allen Hall, 30 N. Dubuque Street, Iowa City, IA 52242, USA}

\author{Daniel M. LaRocca}
\affiliation{University of Iowa Department of Physics and Astronomy, Van Allen Hall, 30 N. Dubuque Street, Iowa City, IA 52242, USA}

\author{Lorella Angelini}
\affiliation{NASA/Goddard Space Flight Center, Greenbelt, MD 20771}




\begin{abstract}

We analyzed data from {\it HaloSat} observations for five fields in the Galactic disk located far away from the Galactic center (135$^{\circ}$~$<$~$l$~$<$~254$^{\circ}$) to understand the nature of soft X-ray energy emission in the Galactic disk. The fields have 14$^{\circ}$ diameter and were selected to contain no significant high-flux X-ray sources. All five HaloSat soft X-ray energy spectra (0.4--7~keV with energy resolution of $<$~100 eV below 1~keV) show a possibility of the presence of unresolved high-temperature plasma in the Galactic disk (UHTPGD) with a temperature of 0.8--1.0 keV and an emission measure of (8--11)$\times10^{-4}~\rm cm^{-6}~pc$ in addition to the soft X-ray diffuse background components mainly studied at higher galactic latitudes (solar wind charge exchange emission, local hot bubble, Milky Way halo emission, and the cosmic X-ray background). This suggests that the UHTPGD is present across the whole Galactic disk. We also observed UHTPGD emission in a region with no bright sources in an {\it XMM-Newton} field contained within one of the {\it HaloSat} fields. The temperature and emission measure are consistent with those measured with {\it HaloSat}. Moreover, the stacked spectra of the X-ray point-like sources and NIR-identified point sources such as stars in the {\it XMM-Newton} field also show a spectral feature similar to the UHTPGD emission. This suggests that the UHTPGD may partly originate from point-like sources such as stars.

\end{abstract}



\section{Introduction} \label{sec:intro}

Diffuse background emission across many wavebands provides important astronomical information. Significant time and effort have been invested to understand the nature of the diffuse backgrounds in the microwave, infrared, and X-ray bands \citep[e.g., ][]{Penzias+1965,Hauser+2001,Shanks+1991}. In the X-ray band, the diffuse background above 2~keV is dominated by the cosmic X-ray background (CXB) that originates mainly from discrete extragalactic sources (e.g., \citet{Hickox+2006}), predominantly AGNs at a wide range of redshifts. The CXB is well described empirically by a powerlaw with a photon index of $\sim$1.4 in the 2-10 keV band (e.g., \citet{Vecchi+1999,Kushino+2002,DeLuca+2004}). However, \citet{McCammon+2002} reveals that the CXB is responsible for only $\sim$40 \% of the total emission in the soft X-ray diffuse background (SXDB) in the $\sim$0.4--1 keV band. This shows the existence of other components. Contributors to the SXDB include: emission due to geocoronal and heliospheric solar wind charge exchange (SWCX) processes \citep[e.g., ][]{Snowden+2004,Ezoe+2010,Yoshitake+2013,Kuntz2019}, hot gas in the solar neighborhood referred to as the Local Hot Bubble (LHB) \citep[e.g., ][]{Liu+2017,Farhang+2019}, and hot plasma in the Milky Way Halo (MWH) \citep[e.g., ][]{Yoshino+2009,Sakai+2014,Nakashima+2018}.

However, the situation changes in the Galactic disk because dense neutral material with $N_{{\rm H}} \sim 10^{21} \rm~cm^{-2}$ or more significantly blocks soft X-ray photons below 1~keV arising from the MWH and CXB. Although it was expected that the SXDB below 1~keV should decrease by at least by $\sim$40\%, the SXDB intensity in the $ROSAT$ $R45$ band ($\sim$0.44--1.0 keV) is only $\sim$20\% lower in the Galactic disk compared to higher Galactic latitudes \citep{McCammon+2002}. This suggests the existence of widely-distributed excess emission that partly compensates for the absorption of the MWH and CXB.

This mystery has been known as the ``M band problem" \citep{McCammon+1990,Cox2005}. The term M band is from an energy band used in the proportional counters in Wisconsin and the Nagoya-Leiden rocket programs; the energy range is almost equal to the $ROSAT$ $R45$ band. The origin of the excess emission has been under debate for a long time. Possible candidates include: hot gas with a temperature of $\sim$3$\times$10$^{6}$ K, dM stars, 
young expanding superbubbles or supernova remnants in low density fields \citep{Nousek+1982,Sanders+1983,Rosner+1981,Cox2005}.

\citet{Masui+2009} conducted a deep pointed observation of a 18$'$$\times$18$'$ blank sky field located in the Galactic disk ($l$, $b$) = (235$^{\circ}$, 0$^{\circ}$) to investigate spectral characteristics of the excess emission. They used {\it Suzaku} which had low and stable non-X-ray background especially in the soft energy band and was thus suitable for study of low-surface brightness soft X-ray sources such as the SXDB \citep{Mitsuda+2007}. They removed point-like X-ray sources to extract a spectrum of the diffuse emission and assumed that the MWH component is completely blocked by dense neutral material associated with the Galactic disk. They revealed an excess component that was well modeled by an unabsorbed, optically-thin collisionally-ionized plasma in thermal equilibrium with a temperature of $\sim$0.8~keV. In this paper, we defined it as unresolved high-temperature plasma in the Galactic disk (UHTPGD). The lack of absorption suggests that the excess X-ray photons come from relatively nearby sources. \citet{Masui+2009} considered a large fraction of the UHTPGD originates from faint dM stars and their model spectrum for spatially unresolved dM stars reproduces the observed features although it cannot compensate entirely the decrease at high Galactic latitudes due to the small scale height. However, the results are only for a single field. It is important to test for the presence of UHTPGD across a wide span of the Galactic disk.

{\it HaloSat} was a small satellite (CubeSat) designed to map the entire soft X-ray sky in order to measure the MWH. It was deployed from the {\it International Space Station} in 2018 \citep{Kaaret+2019} and re-entered Earth's atmosphere in early 2021. {\it HaloSat} had a large field of view (FoV) of near 100 square degrees and a moderate energy resolution of $<$~100 eV below 1 keV \citep{Kaaret+2019,Zajczyk+2019,LaRocca+2020}. It was non-imaging and had no X-ray focusing optics on board. The large FoV and good soft X-ray sensitivity of {\it HaloSat} enable us to study diffuse soft X-ray sources with large angular extent such as the SXDB. 

In this paper, we analyze {\it HaloSat} data to investigate whether UHTPGD emission is found across the extent of the Galactic disk. We selected five {\it HaloSat} fields located in the outer Galactic disk that avoid bright X-ray sources enabling analysis of the diffuse emission. The rest of the paper is organized as follows. Section 2 presents our field selection procedure and data reduction. Section 3, describes our spectral analysis methods and results for the {\it HaloSat} data. In Section 4, we discuss the origin of the detected high temperature plasma and its relation to the spectra of point-like sources measured with {\it XMM-Newton}. Finally, we summarize our results in Section 5.

\begin{table*}[tb]
\footnotesize{
 \caption{{\it HaloSat} and {\it XMM-Newton} observations}
\label{table:obs_info}
  \begin{center}
    \begin{tabular}{llcccc}
\hline\hline
Obs. ID     & Field name            &  \multicolumn{2}{c}{Aim point [deg.]}                           &  Exposure      \\ 
            &                       & (R.A.$_{{\rm J2000}}$, Dec.$_{{\rm J2000}}$) &    ($l$, $b$)    & [ksec]                    \\ \hline
\multicolumn{5}{c}{{\it HaloSat}} \\
HS0017      & HaloSat J0530+3400    & (05 30 33.12, +34 00 18.0)                                &    (174.00, 0.00)  & 32$^\ast$ \\ 
 
HS0019      & HaloSat J0320+5715    & (03 20 05.52, +57 15 43.2)                                &    (142.00, 0.00)  & 23$^\ast$   \\ 
HS0021      & HaloSat J0759-2954    & (07 59 59.28, -29 54 14.4)                               &    (247.00, 0.00) & 28$^\ast$ \\ 
HS0022      & HaloSat J0729-1746    & (07 29 16.08, -17 46 51.6)                               &    (233.00, 0.00)  & 43$^\ast$ \\ 
HS0023      & HaloSat J0700-0430    & (07 00 42.00, -04 30 32.4)                               &    (218.00, 0.00)  & 27$^\ast$ \\ \hline
\multicolumn{5}{c}{{\it XMM-Newton}} \\
0500240101  & V838 Mon              & (07 04 04.85, -03 50 51.1)           & (217.80, 1.05) &  49$^\dagger$ \\ \hline

\hline

    \end{tabular}
\begin{flushleft} 
\footnotesize{
\hspace{3.7cm}$^\ast$ The effective observation exposure after data cleaning of the S14 detector.\\
\hspace{3.7cm}$^\dagger$ Net exposure time of the PN detector after the standard data screening.}
\end{flushleft}
  \end{center}}
\end{table*}

\begin{figure*}
\epsscale{0.75}
\plotone{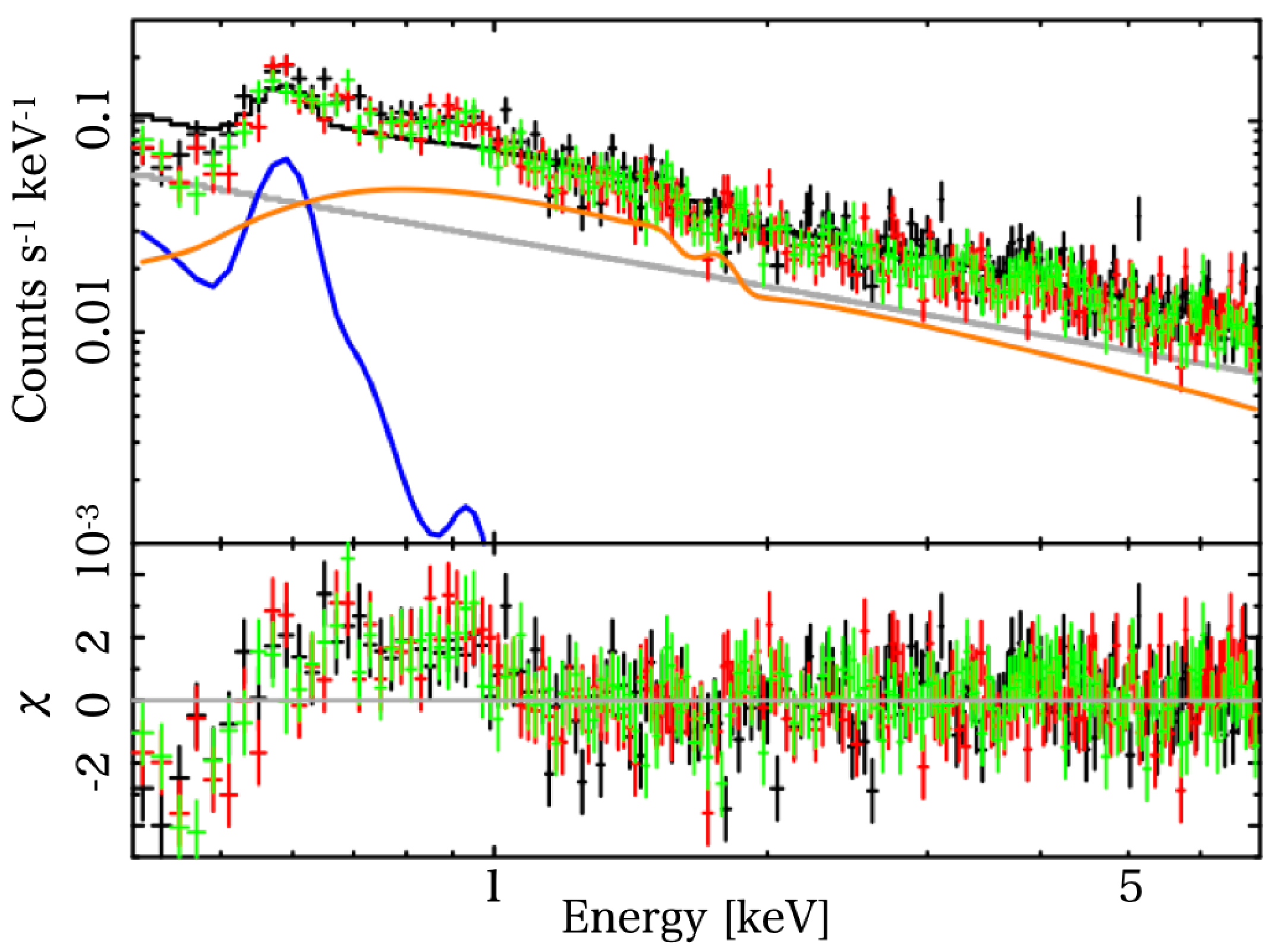}
\caption{Example of the {\it HaloSat} spectral fitting results for the simple model (model0) consisting of the LHB (blue) and CXB (orange). The instrumental background (gray) was modelled with as a powerlaw with a diagonal response matrix. Black, red, and green crosses show the observed data (upper) and residuals subtracting the model from the data divided by error bars (lower) for each detector. For simplicity, the best fit models are shown only for one detector.
}
\label{fig:001701_model0}
\end{figure*}

\section{Field selection and Data reduction} \label{sec:fieldselection_datareduction}

To investigate the X-ray spectral properties of wide-field areas in the Galactic disk, we searched for {\it HaloSat} fields in outer Galactic disk satisfying the following conditions. 

\begin{itemize}
\item aim point with $|b|< 2^{\circ}$ and $|l|>50^{\circ}$,
\item exposure times $>20 \rm \, ksec$ to have good photon statistics,
\item containing no X-ray bright objects in the $ROSAT$ all sky survey \citep{Boller+2016} or the MAXI/SSC all sky catalog \citep{Tomida+2016}
\item containing no extended soft X-ray sources such as the Cygnus superbubble or Vela supernova remnant \citep[e.g., ][]{Bluem+2020,Silich+2020}.
\end{itemize}

The final sample contains a total of 5 fields that are presented in Table \ref{table:obs_info}. Data for the five fields were downloaded from the {\it HaloSat} archive at the HEASARC. Spectra were obtained for each of the three detectors following standard screening on the hard (3--7~keV) and very large event ($>$~7~keV) count rates as described in the {\it HaloSat} analysis document (see details in \url{https://heasarc.gsfc.nasa.gov/docs/halosat/analysis/}). 

We used HEAsoft v6.28 to extract the spectra and XSPEC version 12.11 for the spectral fitting. The response matrix and ancillary response to apply in the spectral  data and as well the response for the particle-induced instrumental background were obtianed from 
the {\it HaloSat} calibration at the HEASARC (\url{https://heasarc.gsfc.nasa.gov/docs/heasarc/caldb/halosat/halosat_caldb.html}). 
Throughout the paper, if not otherwise specified, errors are quoted at the 90\% confidence level and the solar abundance table of \citet{Wilms+2000} was utilized.


\begin{figure*}[h!]
\begin{tabular}{cc}
\begin{minipage}{0.45\hsize}
\hspace{0.2cm}
(a) Model1 (Obs. ID: HS0017)
\begin{center}
\vspace{-0.3cm}
    \includegraphics[width=0.95\linewidth]{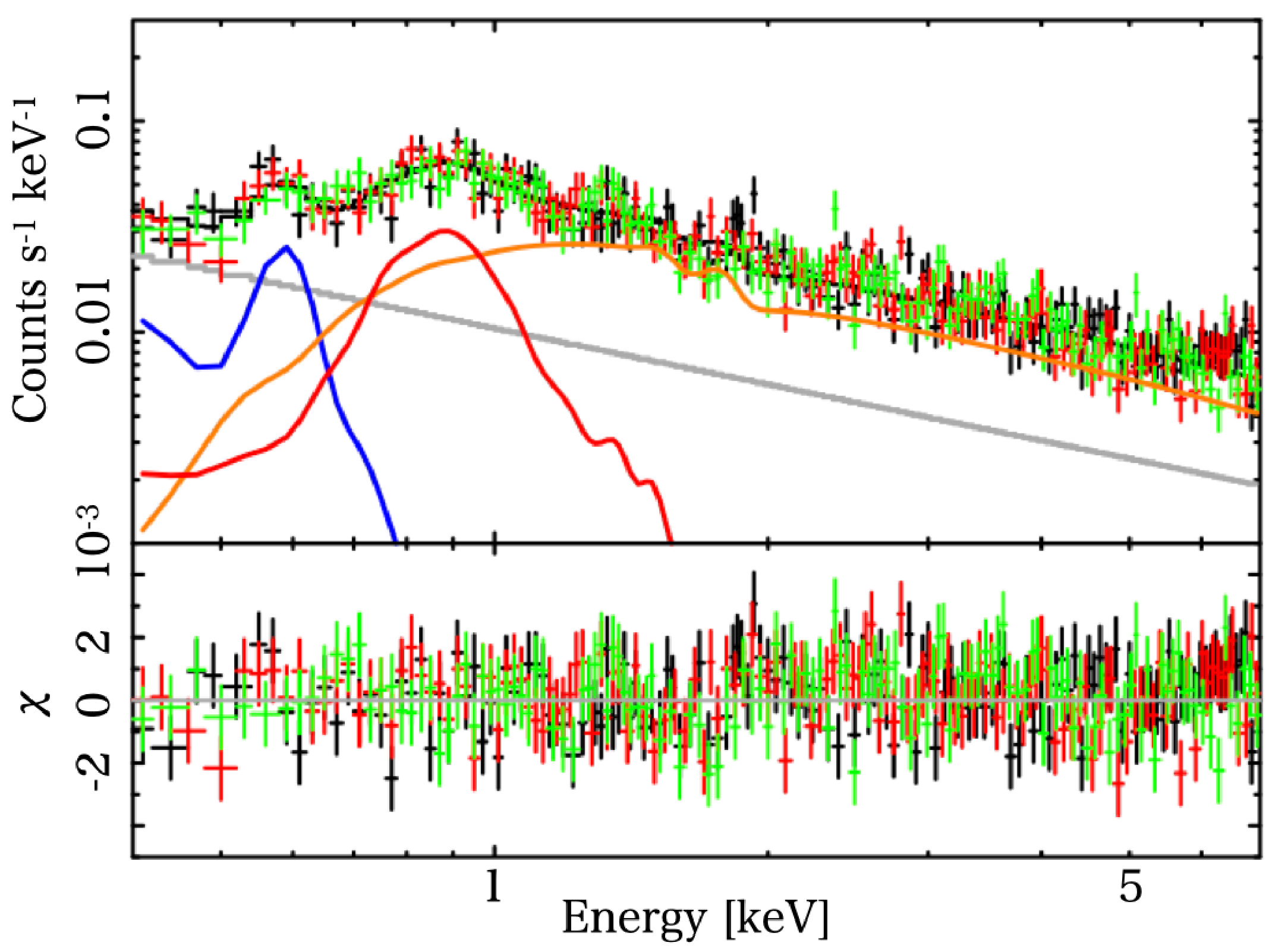}
\end{center}
\end{minipage}
\hspace{0.2cm}
\begin{minipage}{0.45\hsize}
\hspace{0.2cm}
(b) Model2 (Obs. ID: HS0017)
\begin{center}
\vspace{-0.3cm}
    \includegraphics[width=0.95\linewidth]{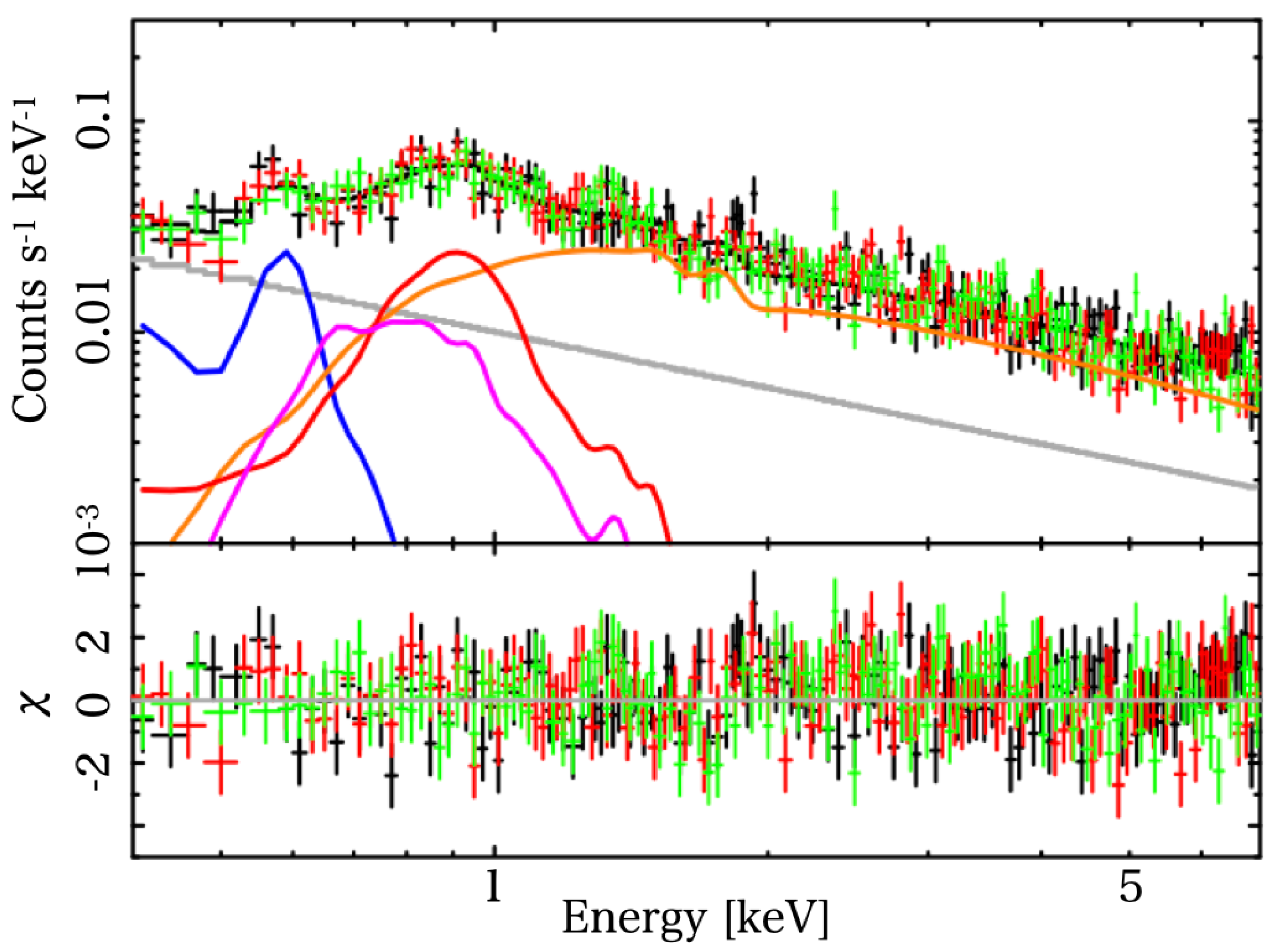}
\end{center}
\end{minipage}
\\
\begin{minipage}{0.45\hsize}
(a) Model1 (Obs. ID: HS0019)
\begin{center}
\vspace{-0.3cm}
    \includegraphics[width=0.95\linewidth]{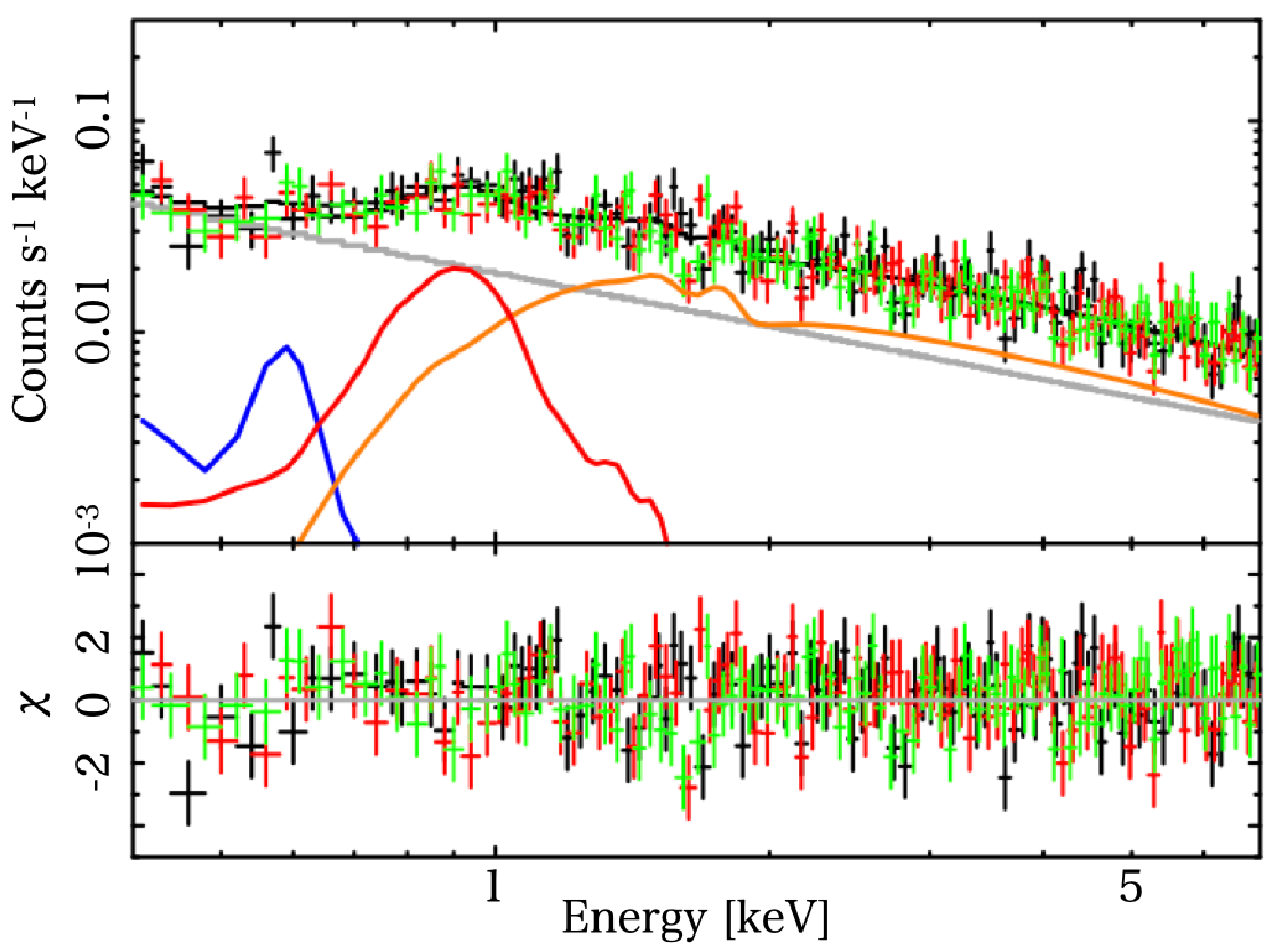}
\end{center}
\end{minipage}
\begin{minipage}{0.45\hsize}
(b) Model2 (Obs. ID: HS0019)
\begin{center}
\vspace{-0.3cm}
    \includegraphics[width=0.95\linewidth]{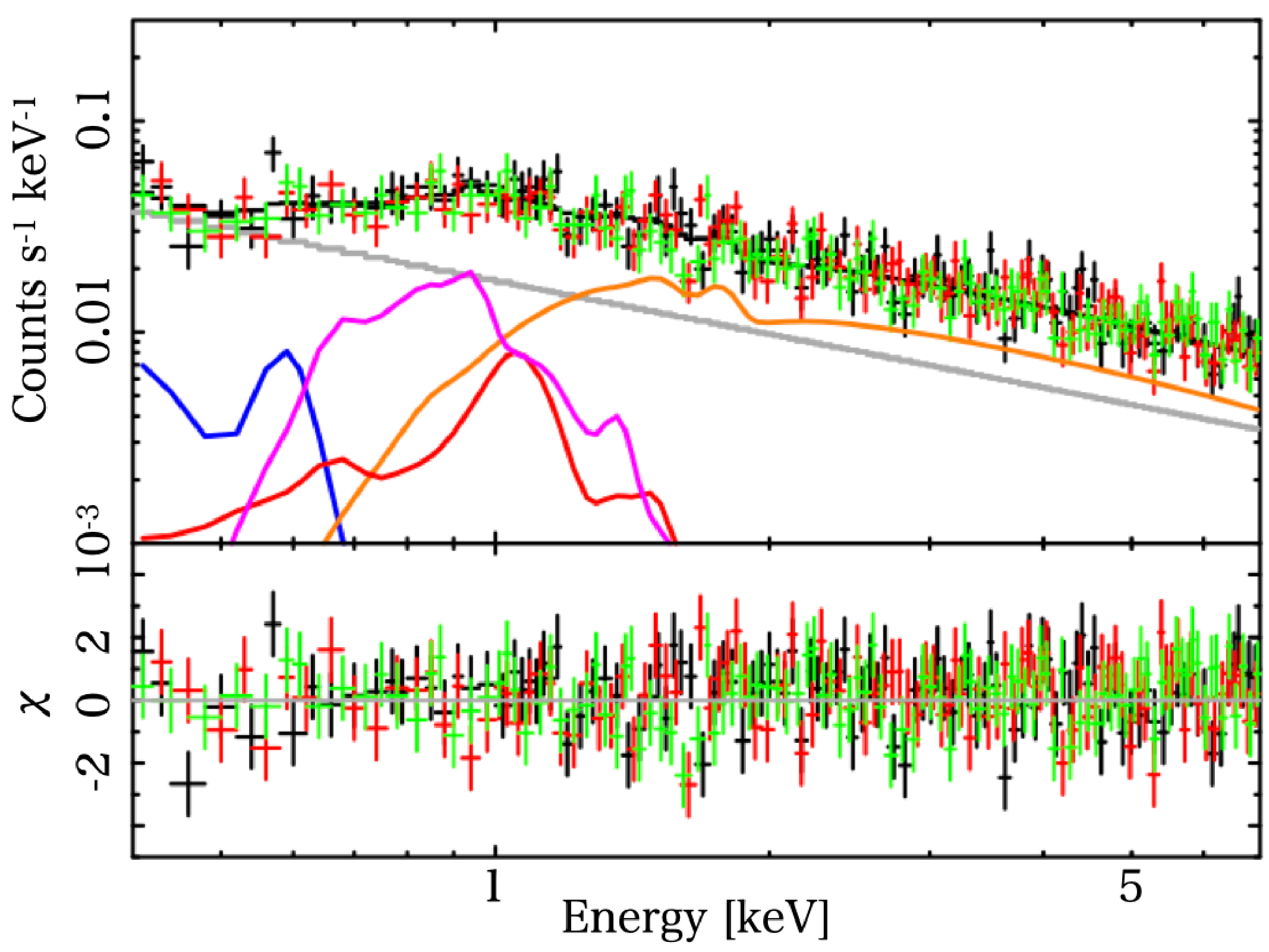}
\end{center}
\end{minipage}
\\
\begin{minipage}{0.45\hsize}
\hspace{0.2cm}
(a) Model1 (Obs. ID: HS0021)
\begin{center}
\vspace{-0.3cm}
    \includegraphics[width=0.95\linewidth]{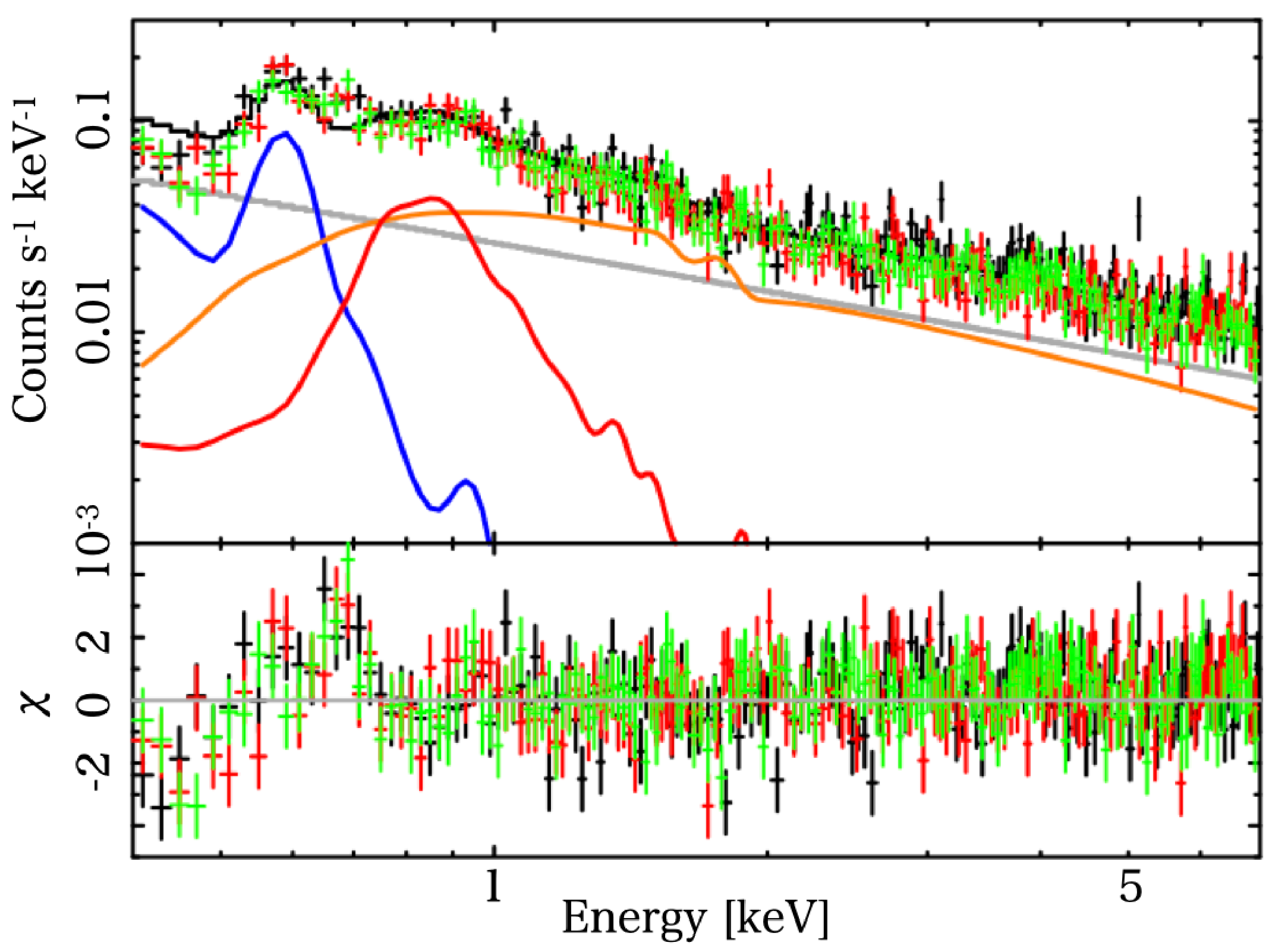}
\end{center}
\end{minipage}
\begin{minipage}{0.45\hsize}
\hspace{0.2cm}
(b) Model2 (Obs. ID: HS0021)
\begin{center}
\vspace{-0.3cm}
    \includegraphics[width=0.95\linewidth]{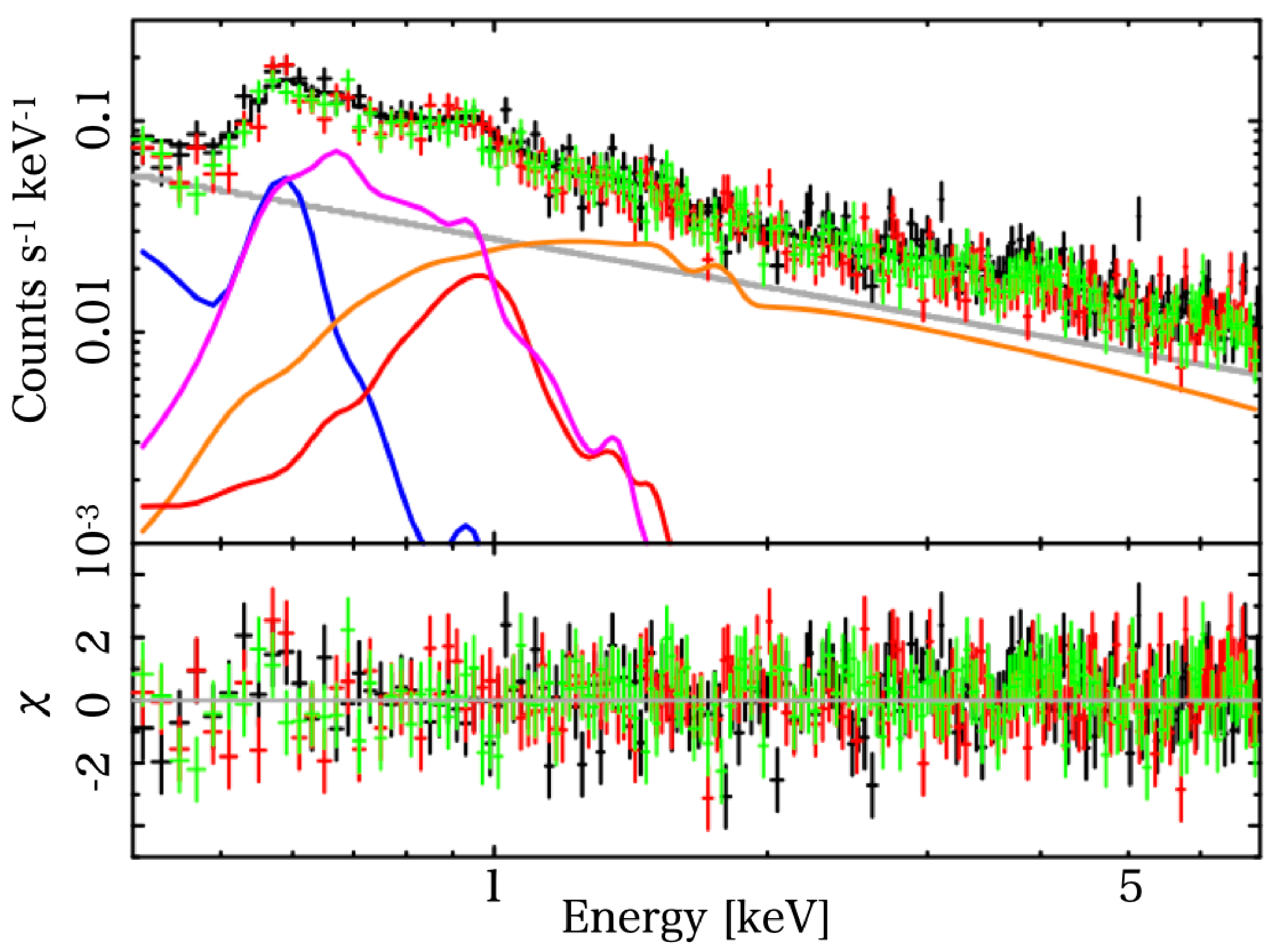}
\end{center}
\end{minipage}
\vspace{0.2cm}
\\
\end{tabular}
  \caption{Observed {\it HaloSat} spectra in the wide fields of the Galactic disk fitted with Model1 (left) and Model2 (right) for the selected five fields. 
  Black, red, and green crosses show the observed data (upper) and residuals subtracting the model from the data divided by error bars (lower) for each detector. 
  Blue and magenta, and orange solid lines correspond to the unabsorbed and absorbed optically thin thermal CIE plasmas corresponding to the LHB and MWH, and the absorbed non-thermal model known as the CXB, respectively.
  Red and gray lines indicate the excess high temperature plasma and instrumental background models, respectively.
  For simplicity, the best fit models are shown only for one detector.}
  \label{fig:spectra}
\end{figure*}
\begin{figure*}[t]
\begin{tabular}{cc}
\begin{minipage}{0.45\hsize}
\hspace{0.2cm}
(a) Model1 (Obs. ID: HS0022)
\begin{center}
\vspace{-0.3cm}
    \includegraphics[width=0.95\linewidth]{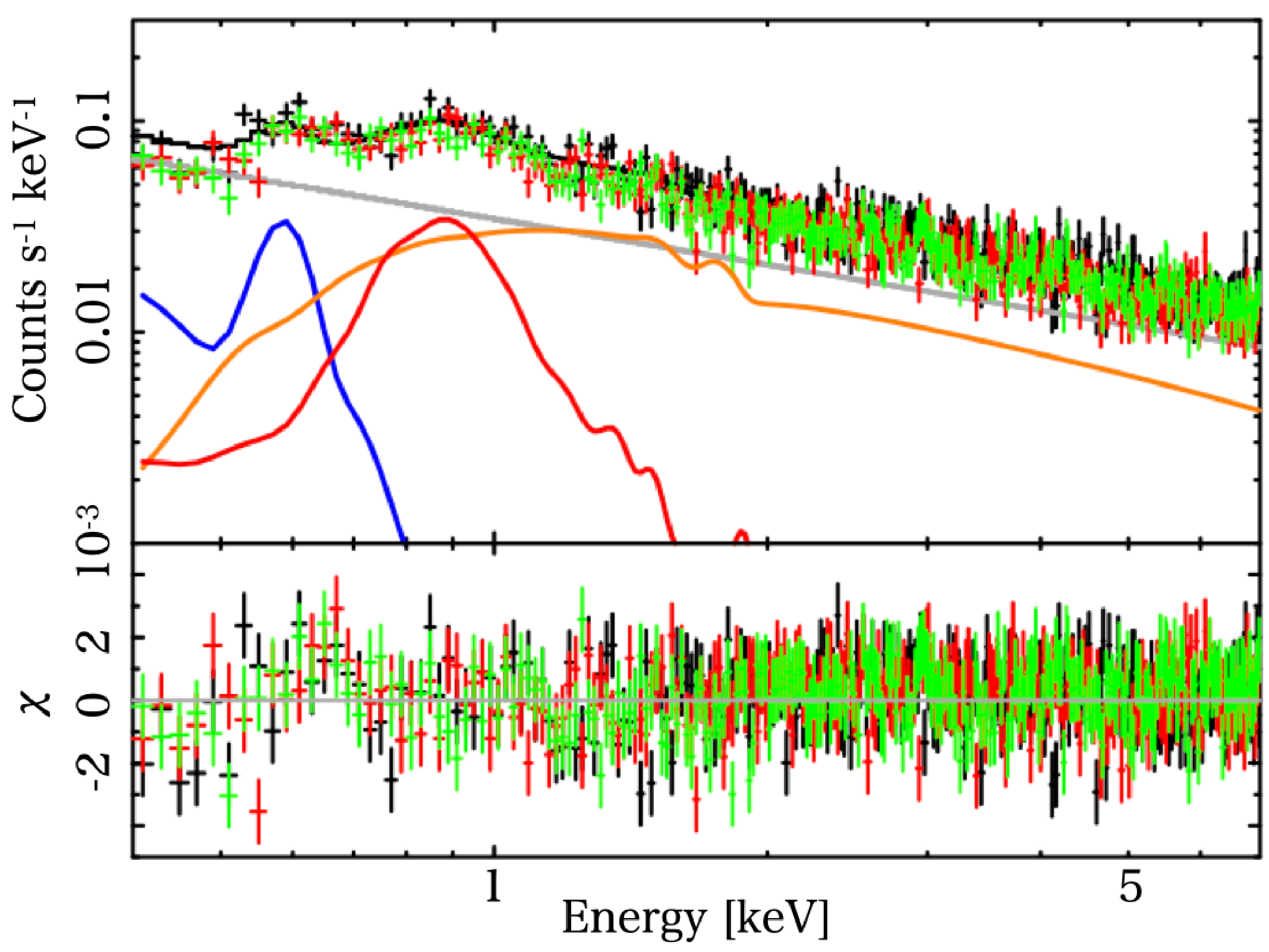}
\end{center}
\end{minipage}
\hspace{0.2cm}
\begin{minipage}{0.45\hsize}
\hspace{0.2cm}
(b) Model2 (Obs. ID: HS0022)
\begin{center}
\vspace{-0.3cm}
    \includegraphics[width=0.95\linewidth]{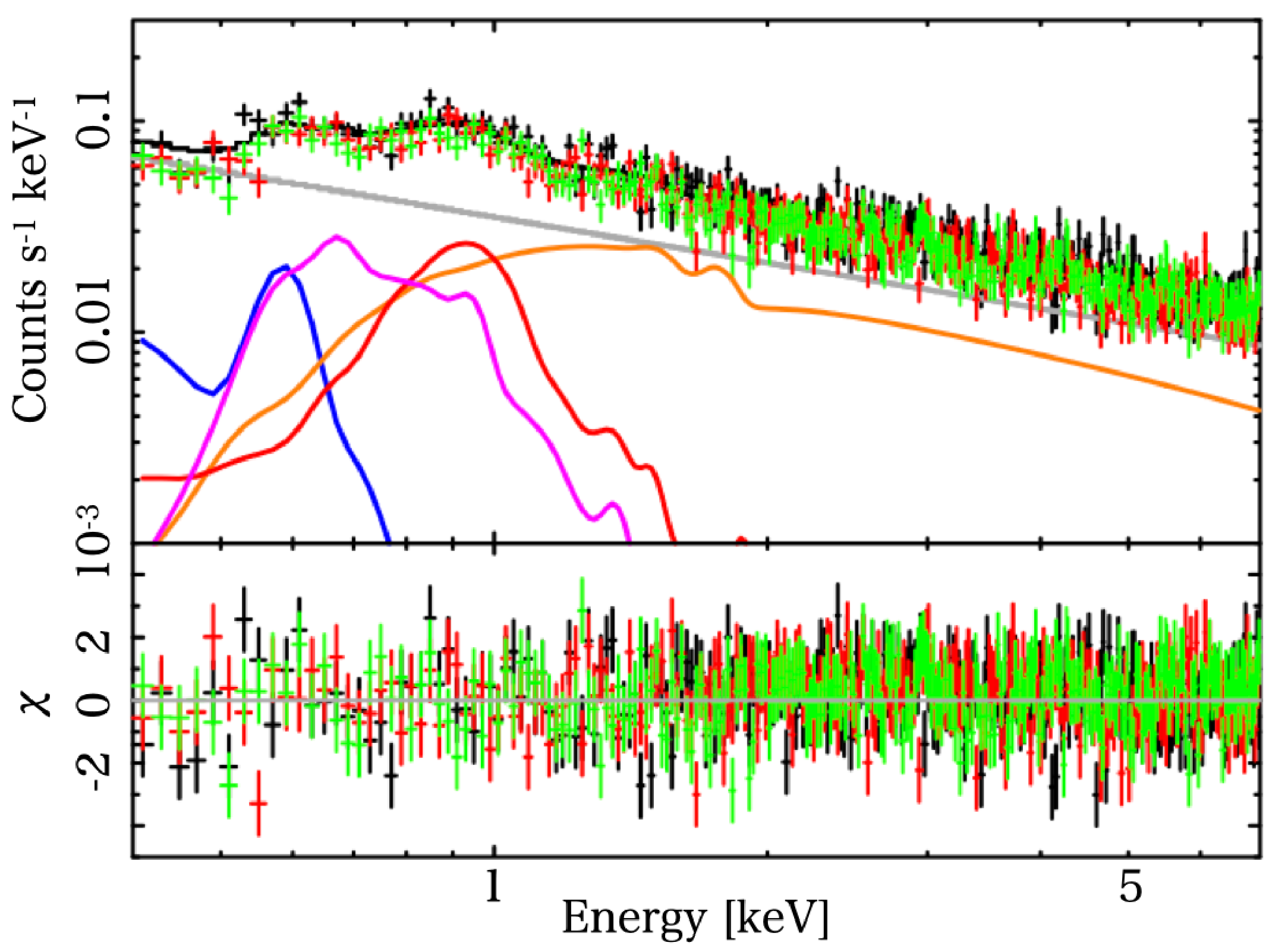}
\end{center}
\end{minipage}
\\
\setcounter{figure}{1}
\begin{minipage}{0.45\hsize}
\hspace{0.2cm}
(a) Model1 (Obs. ID: HS0023)
\begin{center}
\vspace{-0.3cm}
    \includegraphics[width=0.95\linewidth]{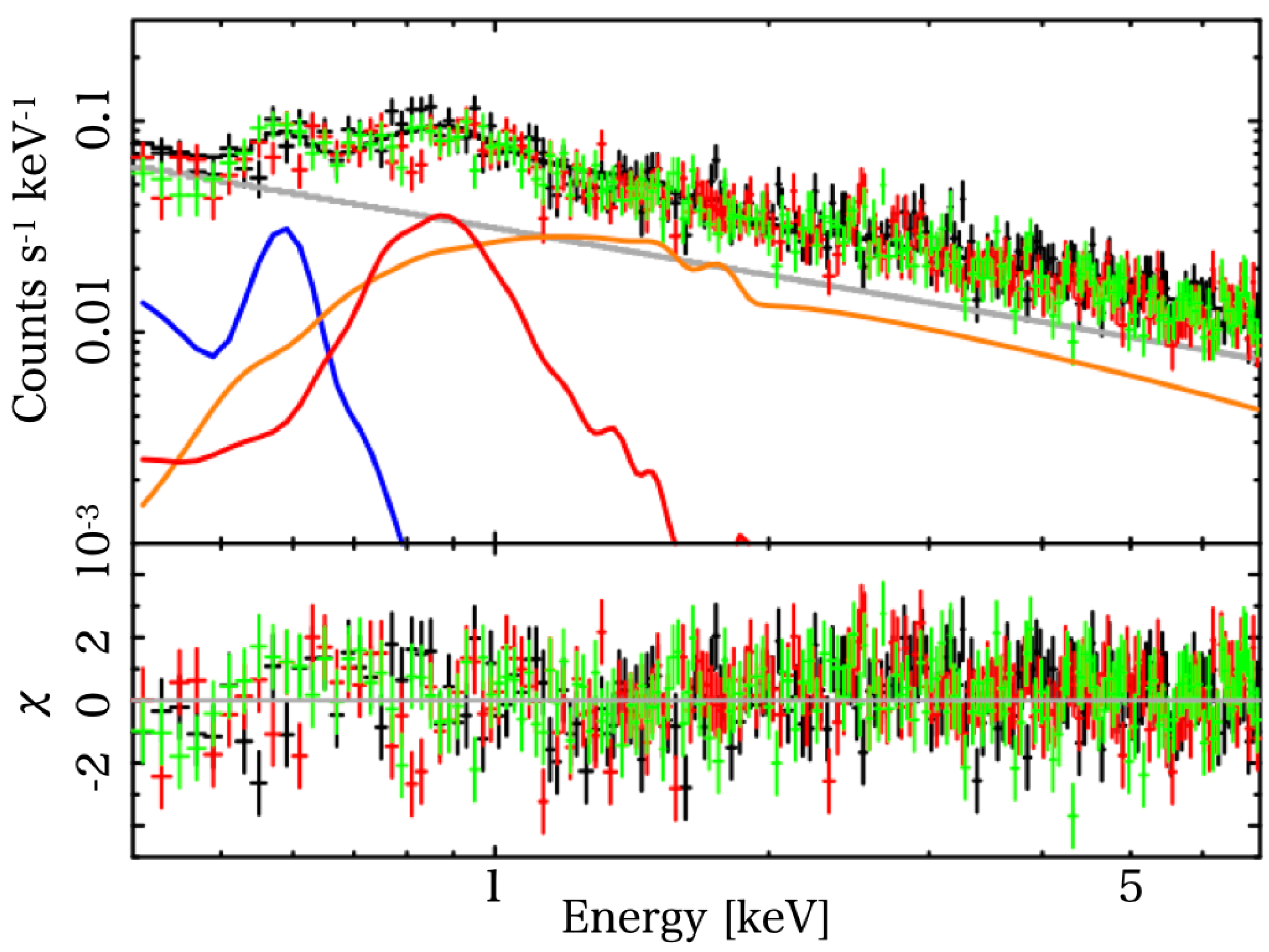}
\end{center}
\end{minipage}
\begin{minipage}{0.45\hsize}
\hspace{0.2cm}
(b) Model2 (Obs. ID: HS0023)
\begin{center}
\vspace{-0.3cm}
    \includegraphics[width=0.95\linewidth]{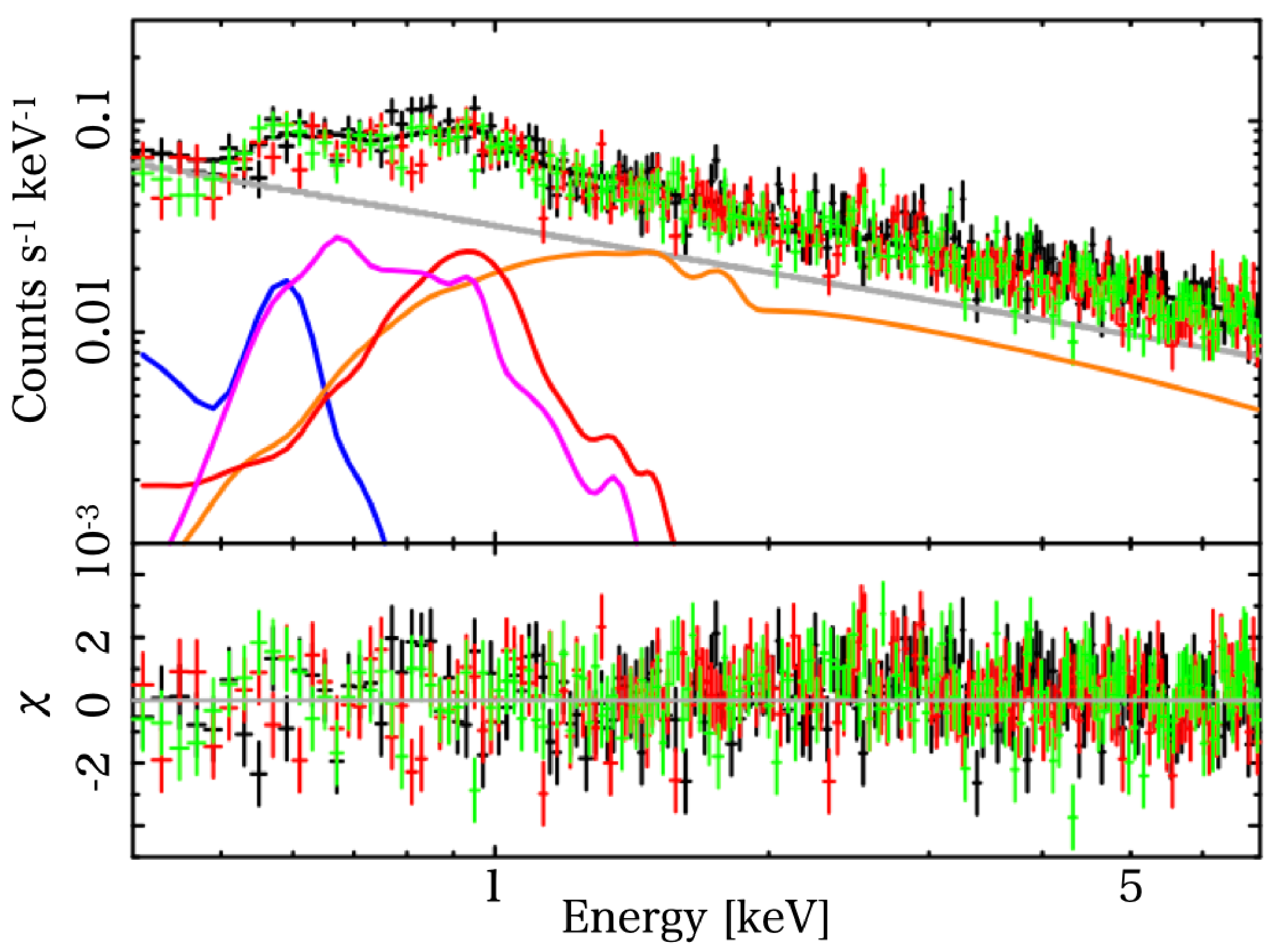}
\end{center}
\end{minipage}
\vspace{0.2cm}
\\
\end{tabular}
  \caption{(Continued)}
  \label{fig:spectra}
\end{figure*}

\section{Analysis and results} \label{sec:analysis_results}

We analyzed the spectra of the five {\it HaloSat} fields. Our spectral fitting procedure is based on previous works modeling the SXDB \citep[e.g., ][]{Mitsuishi+2013}.  Each of the five {\it HaloSat} fields were fitted independently. We fit the spectra for all three detectors simultaneously using common parameters for the astrophysical emission. 

The instrumental background was modeled as a powerlaw and allowed to vary between detectors \citep{Silich+2020,Kaaret+2020}.
The instrumental background photon index and 1-sigma error were calculated for each detector for each observation based on the empirical relation described in \url{https://heasarc.gsfc.nasa.gov/docs/halosat/analysis/back20210209.pdf} obtained using actual data. In our fitting, the instrumental background normalization was a free parameter for each detector while the photon index was allowed to vary within the 1-sigma uncertainty from the empirical relation. In the models below, the instrumental background component is denoted as $powerlaw_\mathrm{IB}$.


As a first step, we adopted a simple model consisting of only the SWCX, LHB and CXB components, i.e.\ assuming that the UHTP emission is negligible and the MWH is completely blocked by the cold dense neutral material of the Galactic disk. We used one unabsorbed optically thin thermal collisionally-ionized equilibrium (CIE) plasma model for the sum of the SWCX and LHB emission and an absorbed powerlaw model for the CXB.
We refer to this as Model0 which is defined in XSPEC as $apec_\mathrm{SWCX+LHB} + tbabs_\mathrm{Galactic} \times powerlaw_\mathrm{CXB} + powerlaw_\mathrm{IB}$, where $tbabs_\mathrm{Galactic}$ gives the absorption along the line-of-sight direction which is mainly associated with the Galactic disk. In this model, the emission measure of the thermal plasma and the Galactic absorption column density were free parameters. We fixed the apec redshift to 0, the apec abundance to 1, and the powerlaw photon index to 1.4 \citep{Kushino+2002}. The plasma temperature was limited to the range $0.097\pm0.014$ keV and the surface brightness of the powerlaw was limited to $(6.38\pm0.64) \times 10^{-8} \rm \, erg \, cm^{-2} \, s^{-1} \, sr^{-1}$ based on previous studies \citep{Kushino+2002,Liu+2017}. 

Fits to this model show significant residuals (see Fig.~\ref{fig:001701_model0}) for all fields except HS0019, especially around 1~keV corresponding to Fe L-shell complex. This suggests that additional thermal components are needed as found in \citet{Masui+2009}. Additionally, residuals in the 0.5--0.7 keV band, corresponding to He- and H- like oxygen emission lines, were found in all of the fields except HS0019, which supports the above scenario. There seems to be no significant residuals for HS0019 because the spectrum below $\sim$1~keV is relatively flat. However, the fitted column density for HS0019 and all the other fields is significantly smaller ($\ll 10^{21} \rm \, cm^{-2}$) than the expected values based on multiwavelength observations \citep{Kalberla+2005,Zhu+2017}, which indicates that the spectrum needs an additional component to make up for soft emission that should be absorbed. Thus, an additional soft component is also needed for HS0019. Note that we considered a heliospheric SWCX scenario for the observed residuals around both 0.5-0.7 and 0.8-1.1 keV by adding multiple gaussian models corresponding to oxygen, neon, and iron emission lines as shown in \citet{Ringuette+2021} and \citet{Huang+2023A}. We found that addition of SWCX emission can reduce the residuals. However the absorption column density remains much lower than the predicted values, suggesting the need for contributions from an additional spectral component. We also considered non-solar abundances for the LHB component as discussed in \citet{McCammon+2002} and reached the same conclusions.The normalization of the instrumental background show typical values and no significant difference is seen in the three detectors. 
%

Therefore, as a next step, we added an unabsorbed thermal plasma component for the UHTPGD emission, that we denote as $apec_\mathrm{UHTPGD}$. We denote this as Model1 which is defined as: $apec_\mathrm{SWCX+LHB} + tbabs_\mathrm{Galactic} \times powerlaw_\mathrm{CXB} + apec_\mathrm{UHTPGD} + powerlaw_\mathrm{IB}$.
This model assumes that all of the hot gas associated with the MWH is blocked completely by the dense neutral material as is the case with \citet{Masui+2009}. Model1 provides a significantly better fit compared to Model0 including only the LHB and CXB components because the residuals around 1~keV were well fitted by the additional thermal plasma component. The UHTPGD components have temperatures of 0.7--0.8 keV and emission measure in the range (8--15)$\times$10$^{-4}$ cm$^{-6}$ pc. The fitted column density became much larger, in the range (0.1--0.7)$\times10^{22}$ [cm$^{-2}$], which is almost consistent with the expected values for each field. The UHTPGD temperature is consistent with that in \citet{Masui+2009} and interestingly, the surface brightness values are consistent with each other within the statistical errors even though the surface brightness in \citet{Masui+2009} is calculated from only one pointed observation of {\it Suzaku} with a FoV of 18$'$$\times$18$'$.
However, residuals around 0.7~keV persist in several fields (HS0021, HS0022, and HS0023). The residuals could alternatively be explained by an LHB component with higher temperature near 0.2~keV if the temperature constraint was not applied. However, the degree of the deviation in the temperature of the LHB in these large fields seems unreasonable considering previous work \citep{Kushino+2002,Liu+2017}. We confirmed that the UHTPGD parameters do not change significantly if a high-temperature LHB is allowed. As another possibility, we tried another fitting model by setting an oxygen abundance free. We confirmed that the results of the temperature and emission measure of the UHTPGD are consistent with each other within the statistical errors even though the resultant parameters are physically unacceptable because the oxygen abundance values are too high over 3.
These results suggest that one more softer thermal component emitting H-like oxygen emissions such as the hot gas associated with the MHW is needed.

\begin{table*}[htb]
  \caption{The results of the model fitting of the five selected {\it HaloSat} fields.
  }
\label{table:best-fit-params_sxdb}
  \begin{center}
    \begin{tabular}{cccccc}
\hline\hline
\multirow{2}{*}{Obs. ID}      & \multirow{2}{*}{Model}  & $N_{\rm H}$                     & $kT_{\rm{UHTPGD}}$  & EM$^\ast_{\rm{UHTPGD}}$           & $\chi^2/$d.o.f  \\  
                              &                         & [$\times 10^{22}~{\rm cm^{-2}}$] & [keV]                & [$\times {\rm 10^{-4}~cm^{-6}~pc}$]   & \\  \cline{3-6}  
                   \hline
\multirow{2}{*}{HS0017} & Model 1 & 0.31$^{+0.11}_{-0.08}$ & 0.77$^{+0.05}_{-0.05}$ & 11$^{+3}_{-2}$ & 452/399\\ \cline{2-6}
                        & Model 2 & 0.42$^{+0.43}_{-0.12}$ & 0.84$^{+0.61}_{-0.08}$ & 9.6$^{+3.2}_{-5.9}$ & 445/397\\ \hline
\multirow{2}{*}{HS0019} & Model 1 & 0.74$^{+0.20}_{-0.18}$ & 0.84$^{+0.11}_{-0.11}$ & 8.2$^{+2.6}_{-2.6}$ & 336/316\\ \cline{2-6}
                        & Model 2 & 0.90$^{+0.23}_{-0.18}$ & 1.3$^{+0.6}_{-0.3}$ & 6.7$^{+7.6}_{-4.3}$    & 321/314\\ \hline
\multirow{2}{*}{HS0021} & Model 1 & 0.098$^{+0.046}_{-0.038}$ & 0.67$^{+0.06}_{-0.07}$ & 15$^{+2}_{-2}$ & 624/493\\ \cline{2-6}
                        & Model 2 & 0.32$^{+0.15}_{-0.09}$ & 0.95$^{+0.31}_{-0.13}$ & 8.4$^{+2.2}_{-2.3}$ & 494/491\\ \hline
\multirow{2}{*}{HS0022} & Model 1 & 0.22$^{+0.07}_{-0.06}$ & 0.77$^{+0.04}_{-0.04}$ & 13$^{+2}_{-2}$ & 790/702\\ \cline{2-6}
                        & Model 2 & 0.37$^{+0.10}_{-0.08}$ & 0.88$^{+0.08}_{-0.06}$ & 11$^{+2}_{-2}$ & 747/700\\ \hline
\multirow{2}{*}{HS0023} & Model 1 & 0.27$^{+0.13}_{-0.10}$ & 0.75$^{+0.06}_{-0.06}$ & 13$^{+3}_{-3}$ & 599/509\\ \cline{2-6}
                        & Model 2 & 0.46$^{+0.17}_{-0.12}$ & 0.88$^{+0.27}_{-0.09}$ & 10$^{+2}_{-5}$ & 569/507\\ \hline
\hline
    \end{tabular}
  \end{center}
\begin{flushleft}
\footnotesize{
\hspace{4.2cm}$\ast$ Emission Measure integrated over the line of sight, 
 i.e. $\rm{EM} = \int n_e n_H dl $.\\
}
\end{flushleft}
\end{table*}

Thus, we added another thermal plasma component, but with absorption equal to that for the CXB, corresponding to the MWH hot gas. The temperature was limited to $0.225\pm0.023$ keV, the abundance was fixed to 0.3 solar, and redshift was fixed to 0, following the results obtained in high-latitude regions \citep{Kaaret+2020}. This model, hereafter Model2, is described in XSPEC as: $apec_\mathrm{SWCX+LHB} + tbabs_\mathrm{Galactic} \times (powerlaw_\mathrm{CXB} + apec_\mathrm{MWH}) + apec_\mathrm{UHTPGD} + powerlaw_\mathrm{IB}$. 

Fitting with Model2 decreases the residuals around 0.7~keV due to the addition of the MHW hot gas. The goodness of fit improves significantly ($>$99.9$\%$ in the F-test) for the fields, HS0021, HS0022, and HS0023, and somewhat ($<$99.9$\%$ in the F-test) for the fields, HS0017 and HS0019. The spectral analysis results for Model1 and Model2 are summarized in Table \ref{table:best-fit-params_sxdb} and the spectral plots and residuals in Figure \ref{fig:spectra}. As for the other models, the normalization of the instrumental background has typical values and no significant difference is seen in the three detectors.

The best fitted column density, temperature, and emission measure of the UHTPGD components are (0.3--0.9)$\times10^{22}$ [cm$^{-2}$], 0.8--1.3 keV, and (7--11)$\times$10$^{-4}$ cm$^{-6}$ pc, respectively. The best fit values of the column density and temperature are systematically higher than those of Model1 and the emission measure decreases as expected with the addition of the MWH component. The emission measure of the LHB is consistent with those of Model1 within the statistical errors except HS0021 and the MWH emission measure is consistent amongst the fields within the statistical errors. For HS0017, all of the results are consistent within the statistical errors between Model1 and Model2. For HS0019, the nominal value of the column density increases significantly but is consistent with the expected value.

In summary, all five HaloSat soft X-ray energy spectra show a possibility of the presence of UHTPGD with a temperature of 0.8--1.0 keV in the best fit models and the parameters are consistent with those found in a smaller field with {\it Suzaku} \citep{Masui+2009}. These results suggest that UHTPGD is distributed across much of the Galactic disk.

\section{A Stellar Origin?} \label{sec:discussion}

To investigate the relation of the observed excess high temperature plasma to point-like X-ray sources including stars, we analyzed {\it XMM-Newton} archival data because {\it XMM-Newton} has relatively large FoV and effective area and good angular resolution which allow us to study properties of both diffuse and point-like sources. We searched the {\it XMM-Newton} archive for observations within our {\it HaloSat} fields with long exposure times. We excluded fields with bright diffuse sources such as clusters of galaxies and supernova remnants. We selected the observation (ObsID: 0500240101) that lies within the {\it HaloSat} field HS0023. We use only the PN data since the PN has the largest effective area. We applied the standard data screening on, e.g., time filtering to remove flaring high-background periods (see \citep[e.g., ][]{Mitsuishi+2013}), resulting in a net exposure time after filtering of 49 ks, long enough to achieve our scientific goals. Information on the observation is shown in Table~\ref{table:obs_info}. 

%
First, we extracted a spectrum of the entire area of the observation with a radius of 12.5 arcmin from the aim point for comparison with the {\it HaloSat} spectra. Fitting with Model0, we found the same type of residuals, e.g. significant residuals around Fe-L complex and H-like oxygen emission lines, as for the {\it HaloSat} spectra. The absorption column density was fixed to be $\sim$0.53$\times$10$^{22}$ cm$^{-2}$ based on the Galactic HI survey \citep{Kalberla+2005}.

\begin{figure*}[h!]
\begin{tabular}{cc}
\hspace{-2.0cm}
\begin{minipage}{1\hsize}
\hspace{0.5cm}
\begin{center}
\vspace{-0.3cm}
    \includegraphics[width=0.5\linewidth]{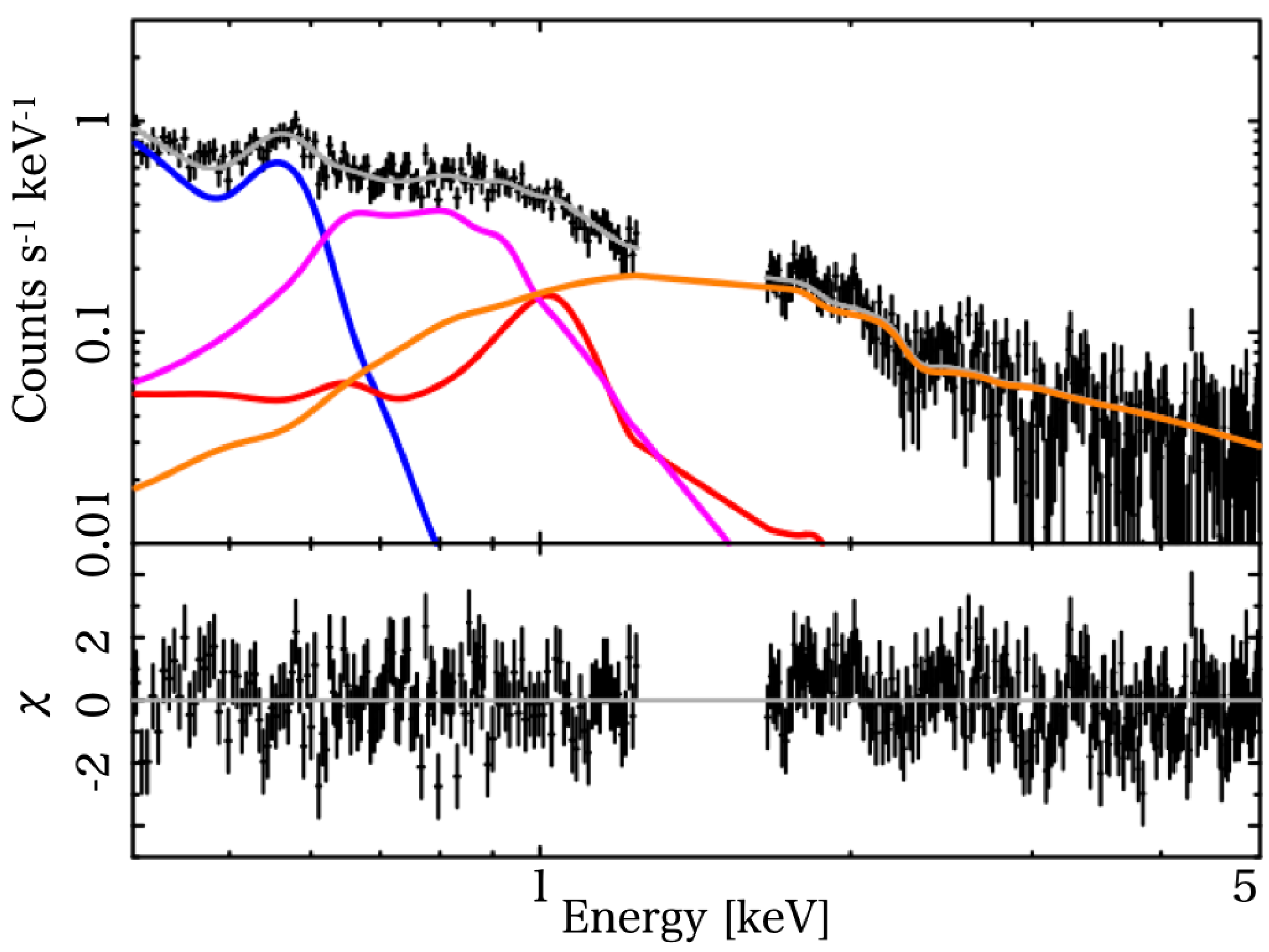}
\end{center}
\end{minipage}
\\
\hspace{-2.0cm}
\begin{minipage}{0.5\hsize}
\begin{center}
    \includegraphics[width=1\linewidth]{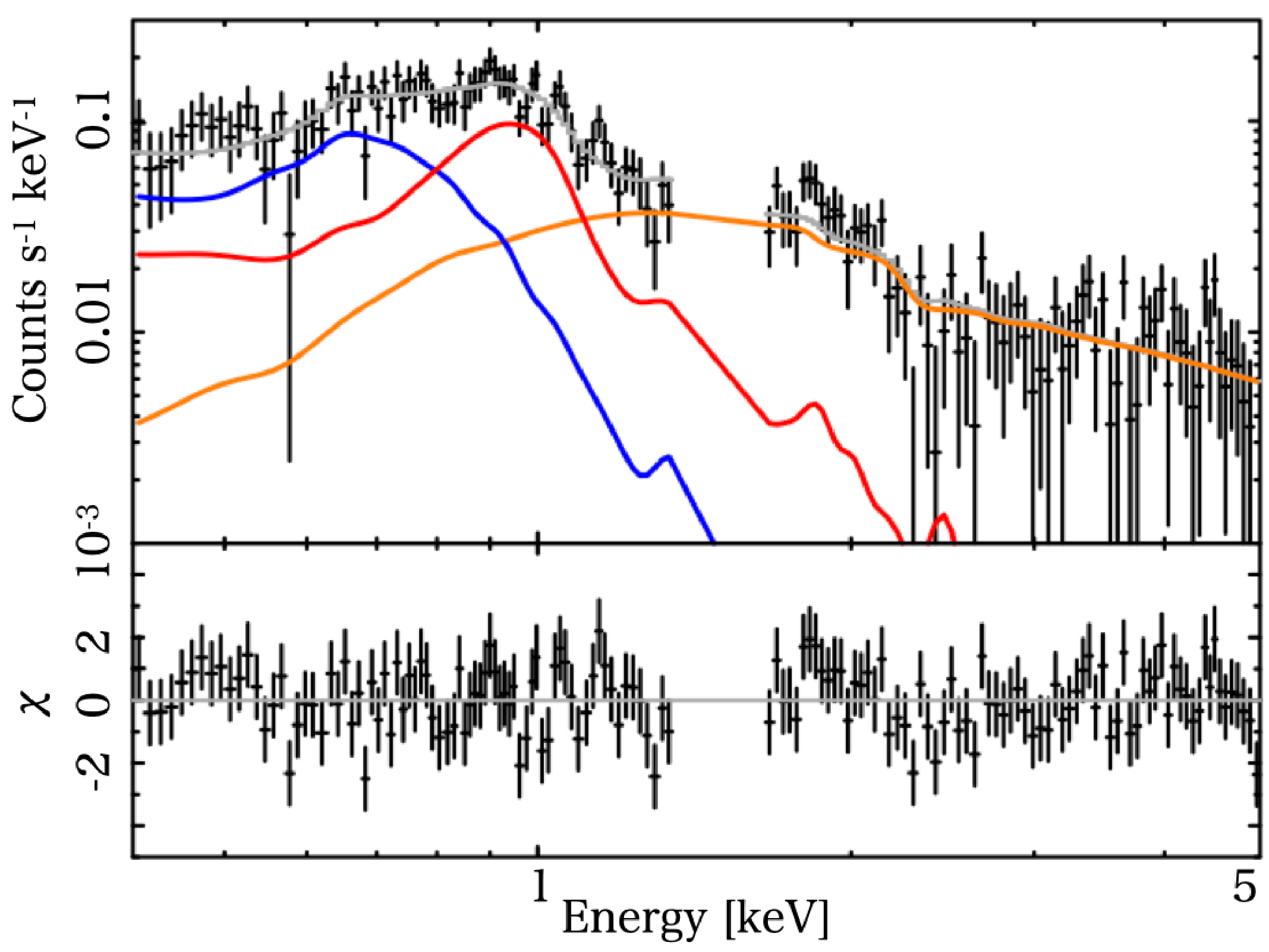}
\end{center}
\end{minipage}
\begin{minipage}{0.5\hsize}
\begin{center}
    \includegraphics[width=1\linewidth]{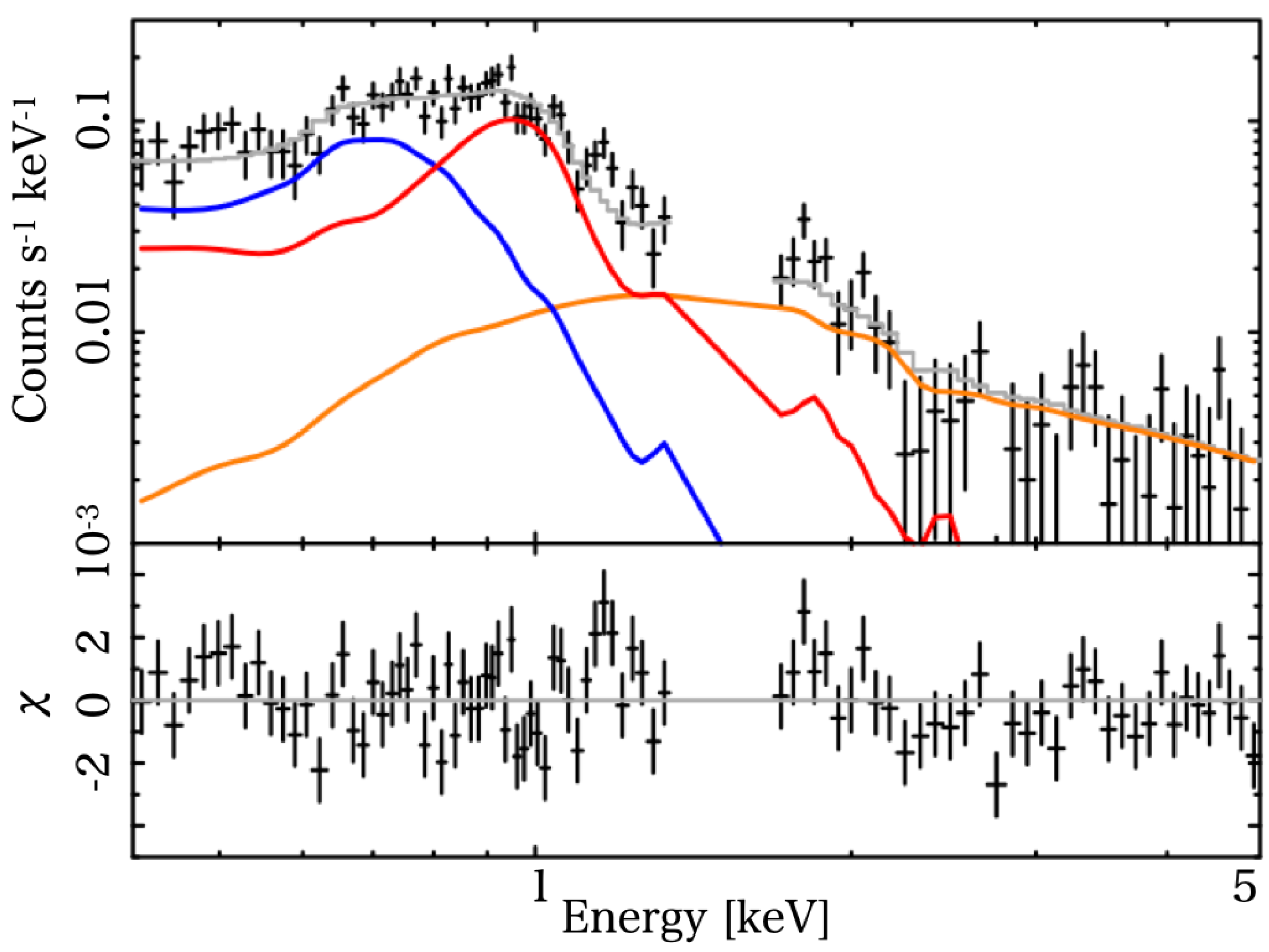}
\end{center}
\end{minipage}
\end{tabular}
  \caption{Spectrum of an {\it XMM-Newton} observation obtained from the entire area (top) and stacked spectra of all of the X-ray point-like sources (bottom left) and cross-matched X-ray point sources with 2MASS point sources (bottom right).
  Top: blue, magenta, red, and orange lines correspond to the SWCX+LHB, MWH, UHTPGD and the CXB, respectively.
  Bottom: blue, red, and orange lines show the low- and high-temperature plasmas and the CXB, respectively.}
  \label{fig:xmmpn_0500240101_spectra}
\end{figure*}

Thus, we fitted the {\it XMM-Newton} spectrum with Model1 and Model2 and found that the fit improves significantly from Model0 to Model1 and then from Model1 and to Model2 similarly to the {\it HaloSat} results. The spectrum fitted with Model2 is shown in Fig.~\ref{fig:xmmpn_0500240101_spectra}. We confirm the presence of high temperature plasma with a temperature of 1.2 keV and emission measure of 12$\times$10$^{-4}$ cm$^{-6}$ pc in Model2. The observed temperature in Model2 is consistent with that for HS0023. The emission measure in Model2 is also consistent, even though the FoV of {\it XMM-Newton} is approximately two orders of magnitude smaller than that of {\it HaloSat}.

To help understand the contribution of resolved point-like sources to the observed high temperature plasma, we investigated a stacked spectrum made of 197 point-like source spectra detected in the {\it XMM-Newton} observation as reported in the catalog \citep[e.g., ][]{Webb+2020}. Source and background spectra were found each point-like source in the catalog. The source regions are circles with radius of 20 arcseconds while the background regions are annuli with radii of 45 and 60 arcseconds, both are centered on the source coordinates tabulated in the catalog. All of the source spectra were summed, the same done for the background, and finally the stacked background spectrum was subtracted from the stacked point-like sources spectrum.

As our first step in analysis of the stacked spectrum, we confirmed that the spectrum was not represented well only by the non-thermal model corresponding to the CXB because of large residuals below $\sim$1 keV including highly-ionized oxygen emission lines and the Fe-L complex were found, suggesting that an additional thermal plasma(s) is needed. We then added unabsorbed optically thin thermal CIE plasma models until no significant (99.9\%) improvement in the goodness of fit was observed. The preferred model includes two plasma components and the CXB. 

The spectrum with the best fit model and the fitting results are indicated in Figure \ref{fig:xmmpn_0500240101_spectra} left and Table \ref{table:best-fit-params_sxdb_xmmnewton}. The plasma temperatures are low ($\sim$0.3~keV) and high ($\sim$1~keV). The latter temperature is consistent with that found for the full observation area while the emission measure is $\sim$50~\%. This suggests that the UHTPGD partly originates from point-like sources. The spectral shape of the low-temperature plasma is similar to that of the MWH. Thus, the contribution of the plasma originated from the point-like sources is included in the emission measure of the MWH of the entire FoV spectral analysis. We confirmed the intensity of the CXB decreases as expected considering the contribution of unresolved point sources.

%

To further study the contribution from stars to UHTPGD as discussed in \citet{Masui+2009}, we used the Two Micron All Sky Survey (2MASS) Point Source Catalog (PSC) \citep{Skrutskie+2006} because most of the 2MASS point sources can be considered to be stars \citep{Cambresy+2001}. Approximately 60\% of the X-ray point-like sources were cross-matched with sources tabulated in the 2MASS PSC with a matching radius of $5\arcsec$. The cross-matched sources have relatively high X-ray fluxes compared to the whole set of point-like X-ray sources in the field. For example, of the 20 brightest X-ray point-like sources tabulated in the catalog, 18 are cross matched in the 2MASS PSC. This may suggest that the observed UHTPGD is mainly due to the contribution of stars.

\begin{figure*}[h!]
\begin{center}
    \includegraphics[width=0.7\linewidth]{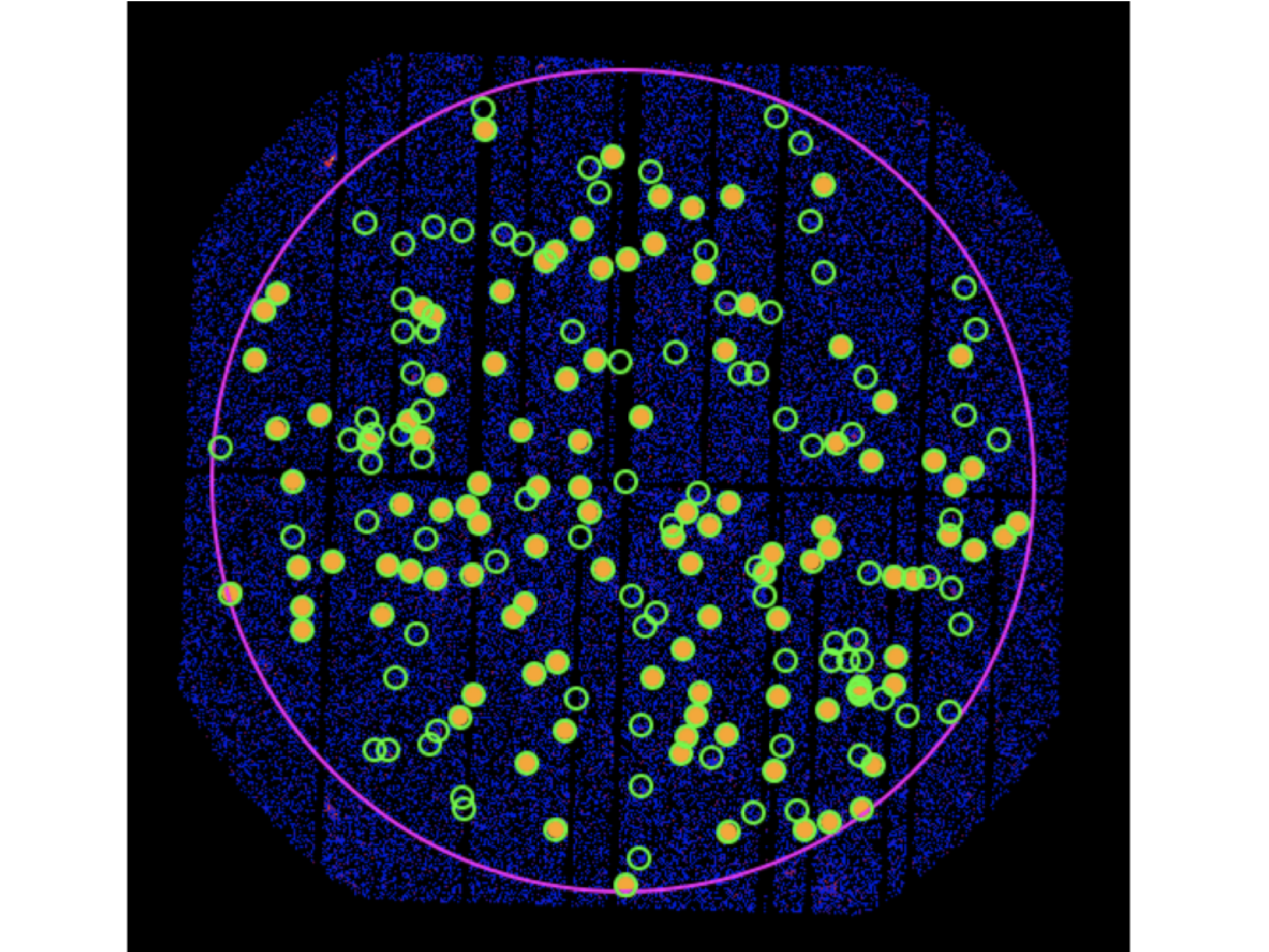}
\end{center}
  \caption{0.4--5.0 keV {\it XMM-Newton} PN image with X-ray point-like sources (green circles).
  The cross-matched X-ray point-like sources with 2MASS (filled orange circle) is also shown.
  Spectral analysis was conducted within a circle with a radius of 12.5 arcmin (magenta).}
  \label{fig:xmmpn_0500240101_image}
\end{figure*}

We extracted the stacked spectrum of X-ray point-like sources cross-matched with the 2MASS PSC and conducted spectral analysis with the same model as described above for the full set of X-ray point-like sources. There are no significant differences between the temperature and emission measure of the two plasmas in the stacked spectra of all versus the NIR-identified point-like sources (see Figure \ref{fig:xmmpn_0500240101_image} bottom and Table \ref{table:best-fit-params_sxdb_xmmnewton}), suggesting that the contribution from stars to the UHTPGD in the entire FoV of the pointed observation is significant.

%

\begin{table*}[h!]
\small{
  \caption{The results of the model fitting for a pointed {\it XMM-Newton} observation.
  }
\label{table:best-fit-params_sxdb_xmmnewton}
  \begin{center}
    \begin{tabular}{cccccc}
\hline\hline
\multirow{2}{*}{} & $kT_{\rm{low}}$ & EM$^\ast_\mathrm{low}$             & $kT_{\rm{high}}$   & EM$^\ast_{\rm{high}}$ & $\chi^2/$d.o.f     \\  
                  & [keV]           & [$\times {\rm 10^{-4}~cm^{-6}~pc}$]& [keV]                 & [$\times {\rm 10^{-4}~cm^{-6}~pc}$] & \\ \hline
\multirow{1}{*}{FoV} & & & 1.2$^{+0.1}_{-0.1}$ & 12$^{+4}_{-3}$                      & 366/317\\ \hline
\multirow{1}{*}{All point source} & 0.28$^{+0.03}_{-0.03}$ & 5.8$^{+0.8}_{-0.9}$ & 0.99$^{+0.08}_{-0.07}$ & 5.8$^{+0.8}_{-0.8}$ & 152/145\\ \hline
\multirow{1}{*}{2MASS point source} & 0.31$^{+0.03}_{-0.03}$ & 4.9$^{+0.6}_{-0.6}$ & 1.0$^{+0.1}_{-0.1}$ & 5.8$^{+0.6}_{-0.6}$ & 130/87\\ \hline\hline
    \end{tabular}
  \end{center}
\begin{flushleft}
\footnotesize{
\hspace{3.0cm}$\ast$ Emission Measure integrated over the line of sight, 
 i.e. $\rm{EM} = \int n_e n_H dl $.\\
}
\end{flushleft}}
\end{table*}

\section{Summary and Conclusions}\label{sec:summary-conclusions}

The composition of the SXDB in the Galactic disk is not well known even though it is essential to our understanding of Galactic structure.
The $ROSAT$ all sky survey revealed excess emission in the R45 band ($\sim$0.44--1.0 keV) widely distributed in the Galactic disk. \citet{Masui+2009} showed that this excess has an energy spectrum consistent with UHTPGD with a temperature of 0.8~keV in one field in the Galactic disk. Using {\it HaloSat}, we investigated the soft X-ray energy spectra of five fields in the outer Galactic disk (135$^{\circ}$~$<$~$l$~$<$~254$^{\circ}$) selected to exclude bright X-ray sources. Our analysis shows that all of the spectra require a thermal plasma component with temperature in the range of 0.8--1.0~keV. Thus, for the first time, we present evidence that UHTPGD is pervasive in the outer Galactic disk. The observed temperatures are not significantly different from the UHTPGD found previously in one field \citep{Masui+2009}, while the emission measure of the UHTPGD observed with {\it HaloSat} varies from (8--11)$\times$10$^{-4}$ cm$^{-6}$ pc and the value is different by a factor of $\sim$1.3. The spectrum of the total area of an {\it XMM-Newton} observation contained within one {\it HaloSat} field shows UHTPGD temperature and emission measure consistent with found with {\it HaloSat}.

To determine whether the observed UHTPGD may originate from stars, we examined the same {\it XMM-Newton} observation and constructed stacked spectra of all X-ray point-like sources detected in the FoV and of the X-ray sources with infrared counterparts in the 2MASS catalog which are predominantly stars. The stacked spectra are well fitted with a model including two thermal plasma components and the CXB. There are no significant differences between the fit results in the two thermal plasma components for the different stacked spectra. The temperature of the highest temperature plasma component is consistent with that observed in spectral analysis for the entire area while the emission measure is $\sim$50~\%. This suggests that the UHTPGD may partly originate from point-like sources such as stars.

\begin{acknowledgments}
This study was financially supported by Grants-in-Aid for Scientific Research
(KAKENHI) of the Japanese Society for the Promotion of Science (JSPS, grant Nos. 19H05609) and by NASA grant, 80NSSC22K0624, "Final Archive of the HaloSat Data".
This research has made use of the SIMBAD database, operated at CDS, Strasbourg, France.
This publication makes use of data products from the Two Micron All Sky Survey, which is a joint project of the University of Massachusetts and the Infrared Processing and Analysis Center/California Institute of Technology, funded by the National Aeronautics and Space Administration and the National Science Foundation.
\end{acknowledgments}

\bibliography{references_sxdb_halosat}

\end{document}